\documentclass{elsart}

\begin{document}
\begin{frontmatter}
\title{A relativistic variant of the Wigner function}
\author{Peter Morgan}
\address{30, Shelley Road, Oxford, OX4 3EB, UK.}
\ead{peter.morgan@philosophy.ox.ac.uk}
\ead[url]{http://users.ox.ac.uk/$\sim$sfop0045}

\begin{abstract}
The conventional Wigner function is inappropriate in a quantum
field theory setting because, as a quasiprobability density over
phase space, it is not manifestly Lorentz covariant.
A manifestly relativistic variant is constructed as a quasiprobability
density over trajectories instead of over phase space.
\end{abstract}

\end{frontmatter}

\newcommand\Half{{\frac{1}{2}}}
\newcommand\Intd{{\mathrm{d}}}
\newcommand\eqN{{\,\stackrel{\mathrm{N}}{=}\,}}
\newcommand\PP[1]{{(\hspace{-.27em}(#1)\hspace{-.27em})}}
\newcommand\PPs[1]{{(\hspace{-.4em}(#1)\hspace{-.4em})}}
\newcommand\RR {{\mathrm{I\hspace{-.1em}R}}}
\newcommand\CC{{{\rm C}\kern -0.5em 
          \vrule width 0.05em height 0.65em depth -0.03em
          \kern 0.45em}}
\newcommand\kT{{{\mathsf k}T}}

\section{Introduction}
The Wigner function is a \emph{quasi}probability density over
phase space (for reviews, see Lee\cite{Lee}, Cohen\cite{Cohen},
or Hillery \textit{et al.}\cite{Hillery}; a \emph{quasi}probability is
generally not positive semi-definite), so in a quantum field theory
setting it is not manifestly Lorentz covariant.
Quasiprobabilities are conceptually interesting partly because
Arthurs-Kelly type interpretations of positive definite Husimi functions
derived by smoothing from Wigner functions are available\cite{AK,BCM,BGL},
but the lack of manifest Lorentz covariance in the quantum field theory
setting makes Wigner functions, in this respect, a poor relation to
Feynman path integrals.

A relativistic variant of the Wigner function that \emph{is} manifestly
Lorentz covariant in a quantum field theory setting is introduced in
section \ref{General}, as a quasiprobability density over
\emph{trajectories} instead of over phase space.
As a formalism, the relativistic variant Wigner function is conceptually
similar to the Feynman integral in its use of trajectories, but is 
conceptually different in its use of quasiprobabilities instead of
phase integrals.
A helpful alternative name for the ``relativistic variant Wigner function''
is the ``trajectory Wigner function'', just because it is a quasiprobability
density over trajectories instead of over phase space.
The emphasis on trajectories here can be compared with the phase space
approach to quantum field theory of Zachos and Curtright\cite{ZC}.

The relativistic variant Wigner function is discussed for the specific
case of the quantized real Klein-Gordon field (called here QKG) in
section \ref{SpecificmQKG}.
It turns out that the relativistic variant Wigner function does not exist
for QKG, prompting the introduction of a modified quantized real
Klein-Gordon field (mQKG).
QKG is a singular limit of mQKG.
The emphasis on QKG as a \emph{field} theory (instead of as a second
quantized particle theory, for example as in \cite[\S 2.5]{Hillery}
and references therein) allows a nonlocality that is present in QKG
to be characterized clearly in terms of the concepts of classical
statistical field theory, and also allows a clear characterization of
both the similarity and the group theoretic difference between quantum
fluctuations and classical thermal fluctuations.
The extension of the relativistic variant Wigner function to other
non-interacting fields is discussed in section \ref{OtherFields}.

\section{The relativistic variant Wigner function in general}
\label{General}
The conventional Wigner function can be presented in non-relativistic
quantum mechanics as the inverse fourier transform of
\begin{equation}
    \left<\psi\right|e^{i\hat x\theta+i\hat p\omega}\left|\psi\right>,
    \quad\mathrm{that\ is,\ as}\quad
    \int\Intd\theta\Intd\omega e^{-ix\theta-ip\omega}
       \left<\psi\right|e^{i\hat x\theta+i\hat p\omega}\left|\psi\right>;
\end{equation}
the Wigner function is a quasiprobability density.
For quantum field theory the Wigner function is the inverse fourier
transform of
\begin{equation}
     \left<\psi\right|e^{i\hat \phi_f+i\hat \pi_g}\left|\psi\right>,
\end{equation}
where
\begin{equation}
     \hat\phi_f=\int_S\hat\phi(x)f(x)\Intd^3x
     \qquad\mathrm{and}\qquad
     \hat\pi_f=\int_S\hat\pi(x)f(x)\Intd^3x
\end{equation}
are smeared field operators on a hypersurface $S$ and $\hat\phi(x)$ and
$\hat\pi(x)$ are operator-valued distributions.
The emphasis on phase space is inappropriate for relativistic quantum field
theory because a phase space approach is not manifestly covariant, so we
will instead consider the inverse fourier transform of the c-number functional
\begin{equation}
  Q_\psi[f]=\left<\psi\right|e^{i\hat\phi_f}\left|\psi\right>
\end{equation}
as the starting point for this paper, where $\hat\phi_f$ is now associated
with all of space-time,
\begin{equation}
  \hat\phi_f=\int\hat\phi(x)f(x)\Intd^4x.
\end{equation}
We can then construct our relativistic variant of the Wigner function as
\begin{equation}\label{TWFDefinition}
  \chi_\psi[w]=\int{\check\mathcal{D}}\!fe^{-i\int f(x)w(x)\Intd^4x} Q_\psi[f]=
            \int{\check\mathcal{D}}\!fe^{-i\int f(x)w(x)\Intd^4x}
                   \left<\psi\right|e^{i\hat\phi_f}\left|\psi\right>
\end{equation}
(the fourier transform functional measure $\check\mathcal{D}\!f$ includes
a factor $(2\pi)^{-1}$ for each of the infinite number of degrees of
freedom that is fourier transformed).
This definition is equally applicable for interacting and for non-interacting
fields.

When $\chi_\psi[w]$ exists, a set of marginal density functionals can be
constructed from it by averaging over degrees of freedom, which includes
probability density functionals that can be constructed from mutually
commuting sets of field observables $\hat\phi_f$.
A paradigm case of a set of mutually commuting field observables is obtained
when we restrict functions $f'$ to be defined on a space-like hyperplane $S$.
Then,
\begin{equation}
    \rho^S_\psi[v|_S]=\int{\check\mathcal{D}}\!f'e^{-i\int f'(x)v(x)\Intd^3x}
            \left<\psi\right|e^{i\hat\phi_{f'}}\left|\psi\right>
\end{equation}
is manifestly a probability density functional, since $\{\hat\phi_{f'}\}$
is effectively a set of classical commuting observables.
Straightforwardly, but heuristically,
\begin{eqnarray}
\rho^S_\psi[v|_S]
    &=& \int{\check\mathcal{D}}\!f'e^{-i\int f'(x)v(x)\Intd^3x}
                 \left<\psi\right|e^{i\hat\phi_{f'}}\left|\psi\right> \cr
    &=& \int{\check\mathcal{D}}\!f'
                 \left<\psi\right|e^{i\int (\hat\phi(x)-v(x))f'(x)\Intd^3x}
                                                    \left|\psi\right>\cr
    & \eqN & \left<\psi\right|\prod_{x\in S}\delta(\hat\phi(x)-v(x))
                                                    \left|\psi\right> \cr
    &\ge& 0,
\end{eqnarray}
where $\eqN$ represents equality up to normalization.

\section{The relativistic variant Wigner function for mQKG}
\label{SpecificmQKG}
For QKG, the algebraic structure of the field is specified by the commutation
relation $[a_g,a_f^\dagger]=(f,g)$, where $a_f^\dagger$ and $a_f$ are smeared
creation and annihilation components of the QKG field,
$\hat\phi_f=a_f^\dagger+a_f$, and $(f,g)$ is a Lorentz invariant positive
semi-definite inner product,
\begin{eqnarray}
  \label{QKGinnerproduct}
  (f,g)&=&\hbar\int\frac{\Intd^3k}{(2\pi)^3}
            \frac{\tilde f^*(k)\tilde g(k)}{ 2\sqrt{k^2+m^2}}\\
       &=&\hbar\int\frac{\Intd^4k}{(2\pi)^4}
            2\pi\delta(k_\mu k^\mu\!-\!m^2)\theta(k_0)
            \tilde f^*(k)\tilde g(k).
\end{eqnarray}
Together, $\hat\phi_f=a_f^\dagger+a_f$, $[a_g,a_f^\dagger]=(f,g)$, and the
action of annihilation operators on the vacuum, $a_g\left|0\right>=0$, fix
all the Wightman functions of the vacuum sector of QKG, so they are
equivalent to other specifications of QKG, including specifications that
require, in part, that $\hat\phi(x)$ satisfies the Klein-Gordon equation.

A 3-dimensional inverse functional fourier transform for the QKG
vacuum does exist,
\begin{eqnarray}\label{QKGkSpace}
  \rho^S_0[v|_S]
     &=&\int{\check\mathcal{D}}\!f'e^{-i\int f'(x)v(x)\Intd^3x}
         \left<0\right|e^{i\hat\phi_{f'}}\left|0\right>\cr
     &=&\int{\check\mathcal{D}}\!f'e^{-i\int f'(x)v(x)\Intd^3x}
         \left<0\right|e^{ia^\dagger_{f'}}e^{-\Half(f',f')}e^{ia_{f'}}
                                         \left|0\right>\cr
     &=&\int{\check\mathcal{D}}\!f'e^{-i\int f'(x)v(x)\Intd^3x}
         e^{-\Half(f',f')}\cr
     &=&\int{\check\mathcal{D}}\!f'e^{-i\int f'(x)v(x)\Intd^3x}
         \exp{\left[-\frac{\hbar}{2}\int\frac{\Intd^3k}{(2\pi)^3}
         \frac{\tilde f^*(k)\tilde g(k)}{ 2\sqrt{k^2+m^2}}\right]}\cr
     & \eqN &\exp\left[-\frac{1}{\hbar}
        \int\frac{\Intd^3k}{(2\pi)^3}
        \tilde v^*(k)\sqrt{k^2+m^2}\tilde v(k)\right],
\end{eqnarray}
where the inversion of the factor $\sqrt{k^2+m^2}$ at the last
line is the standard consequence of the Fourier transform of a
Gaussian.
The fourier-mode kernel $\sqrt{k^2+m^2}$ is nonlocal; $\rho^S_0[v|_S]$
can be converted to a nonlocal real-space description,
\begin{equation}
  \rho^S_0[v|_S] \eqN \exp{\!\left[-\frac{1}{\hbar}
    \int\!\!\!\int\!d^3xd^3y v(x)
    \frac{m^2 K_2(m|x-y|)}{\sqrt{\frac{\pi}{ 2}}|x-y|^2}v(y)\right]},
\end{equation}
where $K_2(m|x-y|)$ is a modified Bessel function.
In terms of the concepts of classical statistical field theory, this
probability density functional characterizes a nonlocality that is
present in QKG.
The dynamical nonlocality is manifest in the appearance of the fourier
mode operator $\tilde f(k)\rightarrow\sqrt{k^2+m^2}\tilde f(k)$, the
nonlocality of which is described by Segal and Goodman\cite{SG}.
This nonlocality is qualitatively the same as the nonlocality of the
heat equation in classical physics, in that it has exponentially reducing
effects at increasing space-like separation, so it should be understood
to be similar to Hegerfeldt-type nonlocality\cite{Hegerfeldt}, rather than
similar to Bell-type nonlocality, which can be a significant effect at
arbitrary space-like separation.
Faster-than-light signals cannot be sent using this nonlocality, as always
in quantum field theory, as long as we insist that measurement operators
commute at space-like separation, which can be understood to be because
the initial states that would allow signals to be sent would require
infinite energy to set up\cite{BY}.
The quantum fluctuations of the QKG field vacuum state described by equation
(\ref{QKGkSpace}) are compared with the thermal fluctuations of a classical
Klein-Gordon field at equilibrium in Appendix \ref{QC}.

Unfortunately, a 4-dimensional inverse functional fourier transform for the
QKG vacuum is \textbf{not} obviously well-defined, because of the appearance
of a delta function in a denominator,
\begin{eqnarray}
  &&\int{\check\mathcal{D}}\!fe^{-i\int f(x)w(x)\Intd^4x}
              \left<0\right|e^{i\hat\phi_f}\left|0\right>\!
           =\!\int{\check\mathcal{D}}\!fe^{-i\int f(x)w(x)\Intd^4x}
              e^{-\Half(f,f)}\cr
           &&\qquad\qquad \eqN \!\exp\left[-\frac{1}{ 2\hbar}
              \int\frac{\Intd^4k}{(2\pi)^4}\frac{\tilde w^*(k)\tilde w(k)}
                  {2\pi\delta(k_\mu k^\mu\!-\!m^2)\theta(k_0)}\right],
\end{eqnarray}
which is undefined if the delta function is understood in a distributional
sense.
To construct a modified quantized real Klein-Gordon field (mQKG), for which
the relativistic variant Wigner function is well-defined, in contrast to QKG,
we replace $\delta(k_\mu k^\mu-m^2)$ by $F(k_\mu k^\mu)$, a positive
semi-definite function (that is, no longer a distribution) of measure 1,
where $F(x)>0$ only if $x\ge 0$.
mQKG\footnote
     {mQKG still satisfies the Wightman axioms. In addition, it conforms
      to the requirements of the cluster decomposition
      theorem\cite[\S 4.4]{Weinberg}, since the algebraic and Lorentzian
      structure and the Hamiltonian of the theory are all unchanged, so
      that the S-matrix satisfies the cluster decomposition principle.}
is defined by the Lorentz invariant inner product
\begin{equation}
  \label{mQKGinnerproduct}
       (f,g)=\hbar\int\frac{\Intd^4k}{(2\pi)^4}
              2\pi F(k_\mu k^\mu)\theta(k_0)\tilde f^*(k)\tilde g(k);
\end{equation}
then for the mQKG vacuum, we obtain a well-defined relativistic variant
Wigner function,
\begin{equation}\chi_0[w] \eqN \exp\left[-\frac{1}{2\hbar}
              \int\frac{\Intd^4k}{(2\pi)^4}\frac{\tilde w^*(k)\tilde w(k)} 
                   {2\pi F(k_\mu k^\mu)\theta(k_0)}\right]\end{equation}
(or, rather, see appendix \ref{GaussianAppendix} and
appendix \ref{RegularizationAppendix} for how it can be made well-defined).

QKG is in this approach a singular, and not obviously well-defined,
limit of mQKG, in which the function $F(k_\mu k^\mu)$ approaches
$\delta(k_\mu k^\mu-m^2)$.
If we regard QKG as only an effective field theory, we can equally
effectively describe a system using mQKG, provided $F(\cdot)$ is as
small off mass-shell as is necessary to reproduce results of experiments.
In general, quantum field theories that are delta-function concentrated
to on mass-shell will be singular limits of quantum field theories like
mQKG, as far as the relativistic variant Wigner function discussed in
this paper is concerned.
Taking trajectories to be sharply confined to a smooth classical
dynamics is not consistent with a measure-theoretic approach to fields
defined on space-time (in contrast, for phase space methods trajectories
\emph{have} to be confined to a single classical dynamics).

We can construct $\chi_\psi[w]$ straightforwardly for arbitrary mQKG
states in a Fock space generated from the vacuum. For the mQKG state
$a_g^\dagger\left|0\right>$, for example, we obtain, by applying the
commutation relations and the action of the annihilation operators on
the vacuum,
\begin{eqnarray}
  \chi_1[w]&=&\int\check\mathcal{D}\!fe^{-i\int f(x)w(x)\Intd^4x}
              \frac{\left<0\right|a_g e^{i\hat\phi_f}a_g^\dagger
                                             \left|0\right>}{(g,g)}\cr
   \noalign{\vskip 2pt}
            &=&\int\check\mathcal{D}\!fe^{-i\int f(x)w(x)\Intd^4x}
              \left[1-\frac{|(g,f)|^2}{(g,g)}\right]
                         e^{-\Half(f,f)}\cr
   \noalign{\vskip 2pt}
            & \eqN &\left[-(g,g)+
               \left|\,\int\limits_{k\in\mathrm{Supp}[F(k_\mu k^\mu)\theta(k_0)]}
                \frac{\Intd^4k}{(2\pi)^4}\tilde g^*(k)\tilde w(k)\right|^2\right]
              \chi_0[w]\cr
   \noalign{\vskip 5pt}
            &=&\left[-(g,g)+|\PP{g,w}|^2\right]\chi_0[w],
\end{eqnarray}
where we have written $\PP{g,w}$ for the neutral inner product.
For a superposition $(v+ua^\dagger_g)\left|0\right>$ of the vacuum
with $a^\dagger_g\left|0\right>$, we obtain
\begin{equation}
    \chi_{01}[w] \eqN \left[-|u|^2(g,g)+|v+u\PP{g,w}|^2\right]\chi_0[w],
\end{equation}
while for the state $a_{g_1}^\dagger a_{g_2}^\dagger\left|0\right>$,
we obtain
\begin{equation}
  \chi_2[w] \eqN 
       \left(\begin{array}{l}
            (g_1,g_2)(g_2,g_1)+(g_1,g_1)(g_2,g_2)\\  \noalign{\vskip 3pt}
            \quad -|\PP{g_1,w}|^2(g_2,g_2)-\PP{g_1,w}\PP{w,g_2}(g_2,g_1)\\
  \noalign{\vskip 3pt}
            \quad\quad -\PP{g_2,w}\PP{w,g_1}(g_1,g_2)-|\PP{g_2,w}|^2(g_1,g_1)\\
  \noalign{\vskip 3pt}
           \quad\quad\quad+|\PP{g_1,w}|^2|\PP{g_2,w}|^2
      \end{array}\right)\chi_0[w];
\end{equation}
when $g_1$ and $g_2$ are orthogonal, $(g_1,g_2)=0$, this reduces to
\begin{equation}
   [-(g_1,g_1)+|\PP{g_1,w}|^2][-(g_2,g_2)+|\PP{g_2,w}|^2]\chi_0[w];
\end{equation}
when $g_1=g_2=g$, it reduces to
\begin{equation}
   [-(2+\sqrt{2})(g,g)+|\PP{g,w}|^2][-(2-\sqrt{2})(g,g)+|\PP{g,w}|^2]\chi_0[w].
\end{equation}

$\chi_1[w]$, $\chi_{01}[w]$, and $\chi_2[w]$ are not positive semi-definite,
as we expect for such a close variant of the Wigner function.
For the coherent state $e^{a_g^\dagger}\left|0\right>$, we obtain
\begin{equation}
   \chi_c[w] \eqN  e^{\PPs{g,w}+\PPs{w,g}}\chi_0[w],
\end{equation}
which is positive semi-definite, as the conventional Wigner function also
is for coherent states.
For arbitrary mixtures of coherent states we obtain positive semi-definite
relativistic variant Wigner functions, but for a superposition
$(c_1e^{a_{g_1}^\dagger}+c_2e^{a_{g_2}^\dagger})\left|0\right>$
of coherent states we obtain
\begin{equation}
   \chi_{sc}[w] \eqN 
         \left(\begin{array}{l}
                 c_1^*c_1e^{\PPs{g_1,w}+\PPs{w,g_1}-(g_1,g_1)}+
                 c_2^*c_2e^{\PPs{g_2,w}+\PPs{w,g_2}-(g_2,g_2)}\,+\\
           \noalign{\vskip 3pt}\qquad
                 c_1^*c_2e^{\PPs{g_1,w}+\PPs{w,g_2}-(g_1,g_2)}+
                 c_2^*c_1e^{\PPs{g_2,w}+\PPs{w,g_1}-(g_2,g_1)}
         \end{array}\right)\chi_0[w],
\end{equation}
which again is not positive semi-definite (unless it is trivial, $g_1=g_2$).
Note that all these relativistic variant Wigner functions are finite order
multinomials in the field $w$ times $\chi_0[w]$, with a closure induced by
the Fock space norm that includes $\chi_c[w]$ and $\chi_{sc}[w]$.

We can also present a thermal state as a positive semi-definite relativistic
Wigner function, invariant under the little group of a unit time-like 4-vector
$\mathcal{T}^\mu$ (see Appendix \ref{ThermalAppendix}),
\begin{eqnarray}
  \chi_T[w] & = & \int{\check\mathcal{D}}\!fe^{-i\int f(x)w(x)\Intd^4x}
             \frac{\mathrm{Tr}\left[e^{-\beta \hat H}e^{i\hat\phi_f}\right]}
              {\mathrm{Tr}\left[e^{-\beta \hat H}\right]}  \cr
\noalign{\vskip 3pt}
      {} & \eqN & \exp\left[-\frac{1}{2\hbar}
              \int \frac{\Intd^4k}{(2\pi)^4}
                  \tanh{\left(\frac{\hbar k_\mu\mathcal{T}^\mu}{2\kT}\right)}
                  \frac{\tilde w^*(k)\tilde w(k)} 
                   {2\pi F(k_\mu k^\mu)\theta(k_0)}\right].\label{ThermalmQKG}
\end{eqnarray}
This presentation of a thermal quantum field state clarifies its
relationship to the vacuum quantum field state in relatively elementary,
albeit also relatively intractable, terms.
A Hilbert space norm is mathematically effective largely because it is a
tight constraint on theory, but the constraint is tight enough that the
vacuum state and a thermal state cannot be presented in the same Hilbert
space.
If we instead use a function space that does not have a separable Hilbert
space structure, the vacuum state and a thermal state can be presented in
a uniform way.
It should also be possible to present the vacuum of an interacting
quantum field in the same function space formalism, by evaluating
equation (\ref{TWFDefinition}).

There are no particles as such in this field approach, but there is a
countable basis for the Fock space, which can lead to the conventional
particle interpretation.
A particle interpretation for quantum field theory is not possible in
general, however, when not only Fock space representations are considered.

\section{Other quantum fields}
\label{OtherFields}
For real interaction-free spin-1 quantum fields, relativistic variant
Wigner functions over real classical spin-1 fields are identical to the
results in section \ref{SpecificmQKG} --- the only change is in the inner
product, between real classical spin-1 test functions, representing the
commutation bracket between smeared operator valued distributions as a c-number.
We can introduce, for example,
\begin{eqnarray}
       [a_g,a_f^\dagger]&=&(f,g)\cr
       &=&\hbar\!\int\frac{\Intd^4k}{(2\pi)^4}
              2\pi F(k_\mu k^\mu)\theta(k_0)\!
               \left[\tilde f_\mu^*(k)\tilde g^\mu(k)-
                \frac{k^\mu\tilde f_\mu^*(k)k_\nu\tilde g^\nu(k)}
                 {k_\alpha k^\alpha}\right]\!.
\end{eqnarray}

It will be interesting to see whether the development given for spin-0
quantum fields can also be extended to spin-$\Half$ quantum fields.
For spin-$\Half$ quantum fields we are of course faced with the
additional difficulty of anticommutation relations, but we have several
choices in considering them:
\begin{list}{$\cdot$}
{\setlength{\itemsep}{2pt}\setlength{\topsep}{0pt}\setlength{\parsep}{0pt}
 \setlength{\parskip}{4pt}\setlength{\partopsep}{0pt}}
\item we can consider fermion fields to be an essentially formal way to
describe a perturbation of boson fields;
\item we can try to develop a bosonization approach in 1+3 dimensions;
\item we can take fermion fields to satisfy commutation relations
instead of anticommutation relations, but modify interactions with
the gauge fields to make the spin-$\Half$ quantum fields stable nonetheless.
\end{list}
Once we represent quantum field theory in terms of quasiprobability densities
over trajectories, we can use different fields as coordinates in our classical
description of trajectories --- so we are free to eliminate some variables in
favour of others --- where the more formal structure of other representations
discourages such freedom.
There may of course be other ways of approaching the question of fermion fields.
If we adopt the last choice above, of commutation relations for spin-$\Half$
quantum fields, there is an obvious vacuum probability density functional
over trajectories of a classical Dirac field $\zeta(x)$,
\begin{equation}
   P_0[\zeta]\eqN\exp{\left[-\frac{1}{\hbar}\!\int\frac{\Intd^4k}{(2\pi)^4}
      \frac{\overline{\tilde \zeta(k)}
       \!\left[k_\mu\gamma^\mu+\sqrt{k_\mu k^\mu}\right]^{-1}
        \!\tilde \zeta(k)}
         {2\pi F(k_\mu k^\mu)\theta(k_0)}\right]}.
\end{equation}
The requirement for anticommutation relations for spin-$\Half$ quantum
fields can be understood to be relative to a requirement for positive
energy, which is only needed for stability when interactions are introduced.
We can ensure stability even if we adopt commutation relations for
spin-$\Half$ quantum fields, provided we introduce interactions in such a way
that the Feynman diagrams in the new description are as they would have been
if we had made the usual choice of anticommutation relations for spin-$\Half$
fields.

It will also be interesting to see whether relativistic variant Wigner
function representations of vacuum states of interacting relativistic
quantum field theories can be constructed as relativistically invariant
modifications of relativistic variant Wigner functions for the vacuum
states of non-interacting quantum fields.
If we can construct a vacuum state $I[w,\zeta]$ of an interacting theory
as a positive semi-definite relativistically invariant modification of a
product $\chi_0[w]P_0[\zeta]$ of non-interacting QKG and spin-$\Half$
vacuums, for example, then other coherent-like states can immediately
be written as $P(w,\zeta)I[w,\zeta]$, where $P(w,\zeta)$ is an
arbitrary positive semi-definite multinomial in components of the
fields $w$ and $\zeta$.

\section{Conclusion}
\label{Conclusion}
We have constructed a relativistic variant of the Wigner function for
quantum field states, which is conceptually preferable to the
conventional Wigner function.
In particular, as a Lorentz covariant formalism, the relativistic variant
Wigner function is an alternative to the Feynman path integral formalism.
We have seen some of the properties of the relativistic variant Wigner
function for the quantized real Klein-Gordon field, or at least we have
for the slightly modified theory, mQKG, and also for other quantum fields.
The distinction between QKG and mQKG is not very great, but the fact that
QKG is singular in terms of the relativistic variant Wigner function is
interesting in itself.

The striking similarity between quantum fluctuations and thermal
fluctuations in a Wigner function formulation of quantum field theory
in terms of fields (whether in the conventional phase space formulation
or in the relativistic variant formulation), and the clarity with which
the difference can be identified, suggests a description of quantum
measurement in which quantum fluctuations are described explicitly.
Equally striking is the nonlocal kernel that in classical statistical
field theory terms is necessary to reproduce the QKG vacuum state.

I am indebted to David Wallace in Oxford for decisive help, given many times,
and to Sheldon Goldstein at Rutgers.
I am also grateful to Chris Isham for comments at a seminar at Imperial
college, and to Stephen Adler and Roderich Tumulka for conversations in
Princeton and at Rutgers, and finally to two anonymous referees.

\appendix
\section{Quantum and classical thermal fluctuations}\label{QC}
In contrast to equation (\ref{QKGkSpace}), the probability density for
the classical Klein-Gordon field at equilibrium on a hyperplane $S$ is
\begin{equation}
  \rho^S_{CKG_E}[v|_S]
      \eqN \exp\left[-\frac{1}{\kT}
        \int\frac{\Intd^3k}{(2\pi)^3}
        {\textstyle\Half}\tilde v^*(k)(k^2+m^2)\tilde v(k)\right],
\end{equation}
where the fourier-mode kernel $(k^2+m^2)$ is local.
Both quantum and classical probability densities restrict non-zero
probability to solutions of the classical Klein-Gordon equation, but
with different densities. The probability density for the QKG vacuum
state is Poincar\'e invariant, in contrast to the Galilean invariance
of the probability density for the classical Klein-Gordon equilibrium
state.

Despite the difference in units and associated functional forms,
Planck's constant of action plays a very similar role in $\chi_0[w]$
to the role played by the Boltzmann energy $\kT$ in a Gibbs
probability density $\exp{[-H[v]/\kT]}$.
Both determine the amplitude of fluctuations.
We have to be careful to remember the difference between the
3-dimensional Galilean symmetry of an equilibrium state and the
(3+1)-dimensional Poincar\'e symmetry of the quantum field theory
vacuum, but the Boltzmann energy and Planck's constant are nonetheless
closely analogous in their effect.

The \emph{difference} between the functional forms of quantum fluctuations
and thermal fluctuations is critical for understanding quantum measurement.
Although quantum fluctuations and thermal fluctuations are both just
fluctuations, we are apparently unable to reduce the ``q-temperature''
of a measurement device below $\hbar$ to reduce the effects of quantum
fluctuations on measurement, whereas we routinely reduce the
temperature of measurement devices to reduce the effects of
thermal fluctuations.
That we \emph{cannot} reduce the q-temperature of a measurement device
\emph{at all} is an empirical principle at the heart of quantum theory,
without which the distinction that quantum theory makes
\emph{in principle} between quantum fluctuations and thermal fluctuations
becomes tendentious:
if we could in practice reduce quantum fluctuations even a little, we
would have to admit the possibility that quantum fluctuations can be reduced
arbitrariliy close to zero, just as we admit for thermal fluctuations.
Even without a present possibility of actually reducing quantum
fluctuations, however, we can nonetheless formulate a description of
quantum measurement in which quantum fluctuations of a measurement
device are explicitly described, just as we explicitly describe thermal
fluctuations of a measurement device.
Note that it is only because quantum fluctuations cannot be eliminated that
measurements using different devices have to be represented by noncommuting
operators in quantum theory.

\section{Inverse functional fourier transform of a positive
   semi-definite Gaussian}
\label{GaussianAppendix}
In a finite dimensional case, it is well-defined to take the inverse fourier
transform of a Gaussian $e^{-q(x)}$, where $q(x)$ is a positive semi-definite
quadratic form, since $q(x)$ splits the space $X\ni x$ into orthogonal
subspaces $X_0, q(x_0)=0,$ and $X_1, q(x_1)>0$.
For the inverse fourier transform we have
\begin{equation}
\int_X e^{-iy.x}e^{-q(x)}=\int_{X_0}e^{-iy_0.x_0}\int_{X_1}e^{-iy_1.x_1}e^{-q(x_1)}
                           \eqN \delta(y_0)e^{-q^{-1}(y_1)},
\end{equation}
where the inverse quadratic form $q^{-1}$ exists on $X_1$.
This simple method extends to mQKG, but, given only a definition of $\delta(x)$
as a distribution, it does not extend to QKG.
If we define $\delta(x)$ as a Colombeau generalized function\cite{Colombeau},
this simple method may possibly extend to QKG.

\section{Regularization of Gaussian integrals}
\label{RegularizationAppendix}
For the functional
\begin{equation}\rho_D[w] \eqN \exp{\left[-\Half\int\frac{\Intd^4k}{(2\pi)^4}
                       \tilde w^*(k)D(k)\tilde w(k)\right]},\end{equation}
where $D(k)$ determines the dynamics of a classical statistical field theory,
the functional integral $\int \mathcal{D}w \rho_D[w]$ only exists in general
if we restrict the range of the functional integration to functions that are
smooth below a chosen scale (for a straightforward discussion,
see \cite[\S8.1 and Appendix L]{Binney}).
This integral must be finite for us to regard $\rho_D[w]$ as a probability
density functional (implicitly assuming normalization), as must the moments
of the distribution.
A simple way to ensure finiteness is to introduce a wave number cutoff,
$|k|<\Lambda$, for some Euclidean metric on $k$.

The Gaussian model of classical statistical field theory takes $D(k)=|k|^2+m^2$,
which progressively reduces the probability of higher frequency components of
$\tilde w(k)$ (but not sufficiently to give a finite functional integral when
$\Lambda\rightarrow\infty$ except for one dimensional systems). In contrast,
mQKG takes $D(k)=[2\pi F(k_\mu k^\mu)\theta(k_0)]^{-1}$, where $F(k_\mu k^\mu)$
has support, say, only for $m^2<k_\mu k^\mu<m^2+\delta$, near the hyperboloid
$k_\mu k^\mu=m^2$, so the functional integrals of mQKG are already constrained
to functions $\tilde w(k)$ having support only where $k_\mu$ is in the support of
$F(k_\mu k^\mu)\theta(k_0)$. mQKG can be treated
in the same way as the well-understood Gaussian model, and the functional integral
$\int \mathcal{D}w \rho_D[w]$ and the moments of the probability density are all
finite for $|k|<\Lambda$ (but not for $\Lambda\rightarrow\infty$). One difficulty
is that this regularization breaks Lorentz invariance, but this is always a
difficulty for simple regularizations of relativistic quantum field theory.

\section{Thermal state characteristic function}
\label{ThermalAppendix}
For the characteristic function of a thermal state of a simple harmonic
oscillator, we have:
\begin{eqnarray}
    Q_T[z] & = & \frac{\mathrm{Tr}{\left[e^{-\lambda a^\dagger a}
                       e^{i(a^\dagger z^*+az)}\right]}}
                  {\mathrm{Tr}{\left[e^{-\lambda a^\dagger a}\right]}}\cr
           & = & \frac{\mathrm{Tr}{\left[e^{-\lambda a^\dagger a}
                       e^{ia^\dagger z^*}e^{iaz}\right]}e^{-\Half \alpha|z|^2}}
                  {\mathrm{Tr}{\left[e^{-\lambda a^\dagger a}\right]}};
\end{eqnarray}
where we will suppose that $[a,a^\dagger]=\alpha$. Then
\begin{equation}
     \mathrm{Tr}{\left[e^{-\lambda a^\dagger a}\right]}=
                 \frac{1}{1-e^{-\lambda\alpha}}\ ,
\end{equation}
and
\begin{eqnarray}
   \hspace{-12pt}
   \mathrm{Tr}{\left[e^{-\lambda a^\dagger a}e^{ia^\dagger z^*}e^{iaz}\right]} & = &
        \mathrm{Tr}\!\left[\left(1-\frac{|z|^2 a^\dagger a}{1!^2}+
               \frac{|z|^4 a^{\dagger 2} a^2}{2!^2}-
               \frac{|z|^6 a^{\dagger 3} a^3}{3!^2} + ...\right)
                e^{-\lambda a^\dagger a}\right]\cr
 & \hspace{-16em}  = & \hspace{-8em} 
        \mathrm{Tr}\!\left[\left(1-\frac{|z|^2 a^\dagger a}{1^2}
               \left[1-\frac{|z|^2 (a^\dagger a-\alpha)}{2^2}
               \left[1-\frac{|z|^2 (a^\dagger a-2\alpha)}
                {3^2}\Bigg[...\Bigg]\right]\right]\right)
                e^{-\lambda a^\dagger a}\right]\cr
 & \hspace{-16em}  = & \hspace{-8em} 
       \left(1+\frac{|z|^2}{1^2}\frac{\mathrm{d\ }}{\mathrm{d}\lambda}
               \left[1+\frac{|z|^2}{2^2} (\frac{\mathrm{d\ }}{\mathrm{d}\lambda}+\alpha)
               \left[1+\frac{|z|^2}{3^2} (\frac{\mathrm{d\ }}{\mathrm{d}\lambda}+2\alpha)
                           \Bigg[...\Bigg]\right]\right]\right)
               \mathrm{Tr}{\left[e^{-\lambda a^\dagger a}\right]}\cr
 & \hspace{-16em}  = & \hspace{-8em} 
      \frac{1}{1-e^{-\lambda\alpha}}\left[1
             -\frac{\alpha |z|^2e^{-\lambda\alpha}}{1!(1-e^{-\lambda\alpha})}
             +\frac{\alpha^2 |z|^4e^{-2\lambda\alpha}}{2!(1-e^{-\lambda\alpha})^2}
             -\frac{\alpha^3 |z|^6e^{-3\lambda\alpha}}{3!(1-e^{-\lambda\alpha})^3}
             +...\right]\cr
 & = & \frac{1}{1-e^{-\lambda\alpha}}
    \exp{\left[-\frac{\alpha |z|^2 e^{-\lambda\alpha}}{1-e^{-\lambda\alpha}}\right]},
\end{eqnarray}
where we have used
\begin{equation}
  \mathrm{and}\qquad
  \begin{array}{l}
     a^{\dagger n}a^n=a^\dagger a (a^\dagger a-\alpha)
       (a^\dagger a-2\alpha)...(a^\dagger a-(n-1)\alpha)\\
     \noalign{\vskip 2pt}
     \mathrm{Tr}[(a^\dagger a)^n e^{-\lambda a^\dagger a}] =
        (-1)^n\frac{\mathrm{d}^n}{\mathrm{d}\lambda^n}
         \mathrm{Tr}[e^{-\lambda a^\dagger a}],
  \end{array}
\end{equation}
so that
\begin{equation}
  Q_T[z] = \exp{\left[-\frac{\alpha |z|^2}{2\tanh{\lambda\alpha}}\right]}.
\end{equation}
For mQKG, we take
\begin{equation}
  \hat H=\int\frac{a^\dagger(k) a(k) k_\mu \mathcal{T}^\mu}
      {2\pi F(k_\mu k^\mu)\theta(k_0)}
         \frac{\Intd^4k}{(2\pi)^4}
\end{equation}
to obtain equation \ref{ThermalmQKG}.

\newcommand\JournalName[1]{\textsl{#1}}
\newcommand\JournalVolume[1]{\textbf{#1}}
\newcommand\BookName[1]{\textit{#1}}


\begin{thebibliography}{00}
\bibitem{Lee}
   {Lee, H. W., \JournalName{Phys. Rep.} \JournalVolume{259}, 147(1995).}
\bibitem{Cohen}
   {Cohen, L., \JournalName{Proc. IEEE} \JournalVolume{77}, 941(1989).}
\bibitem{Hillery}
   {Hillery, M., O'Connell, R. F., Scully, M. O., and Wigner, E. P.,
       \JournalName{Phys. Rep.} \JournalVolume{106}, 121(1984).}
\bibitem{AK}
   {Arthurs, E. and Kelly, J. L., Jr.,
       \JournalName{Bell. Syst. Tech. J.} \JournalVolume{44}, 725(1965).}
\bibitem{BCM}
   {Braunstein, S. L., Caves, C. M., and Milburn, G. J.,
       \JournalName{Phys. Rev.} \JournalVolume{A 43}, 1153(1991).}
\bibitem{BGL}
   {Busch, P., Grabowski, M., and Lahti, P. J.,
       \BookName{Operational Quantum Physics},
          Springer Lecture Notes in Physics \JournalVolume{m 31}(1995).}
\bibitem{ZC}
   {Zachos, C. and Curtright, T.,
       \JournalName{Prog. Theor. Phys. Suppl.}
       \JournalVolume{135}, 244(1999);\newline
       \verb=http://arxiv.org/abs/hep-th/9903254=}
\bibitem{SG}
   {Segal, I. E. and Goodman, R. W.,
       \JournalName{J. Math. and Mech.} \JournalVolume{14}, 629(1965).}
\bibitem{BY}
   {Buchholz, D. and Yngvason, J.,
       \JournalName{Phys. Rev. Lett.} \JournalVolume{73}, 613(1994).}
\bibitem{Hegerfeldt}
   {Hegerfeldt, G. C., in \BookName{Irreversibility and causality},
       Bohm, A., Doebner, H.-D., Kielanowski, P.(eds.),
       Springer Lecture Notes in Physics
       \JournalVolume{504}, 238(1998);\newline
       \verb=http://arxiv.org/abs/quant-ph/9806036=}
\bibitem{Weinberg}
   {Weinberg, S., \BookName{The Quantum Theory of Fields}
     (Cambridge University Press, Cambridge, 1995), Volume I.}
\bibitem{Colombeau}
   {Colombeau, J. F.,
        \JournalName{Bull. A. M. S.} \JournalVolume{23}, 251(1990);\newline
    Colombeau, J. F., \BookName{Multiplication of distributions},
        Springer Lecture Notes in Mathematics \JournalVolume{1532}(1992).}
\bibitem{Binney}
   {Binney, J. J., Fisher, A. J., Dowrick, N. J., and Newman, M. E. J.,
        \BookName{The Theory of Critical Phenomena}
         (Oxford University Press, Oxford, 1992).}
\end{thebibliography}
\end{document}